\documentclass[%
 aip,
rsi,
 amsmath,amssymb,
 reprint,%
]{revtex4-2}

\usepackage[caption=false]{subfig}
\usepackage{graphicx}
\usepackage{dcolumn}
\usepackage{bm}
\usepackage{textcomp, gensymb}

\usepackage{ulem}
\usepackage[utf8]{inputenc}
\usepackage[T1]{fontenc}
\usepackage{mathptmx}
\usepackage{etoolbox}
\usepackage{xcolor,soul}

\makeatletter
\def\@email#1#2{%
 \endgroup
 \patchcmd{\titleblock@produce}
 {\frontmatter@RRAPformat}.
 {\frontmatter@RRAPformat{\produce@RRAP{*#1\href{mailto:#2}{#2}}}\frontmatter@RRAPformat}
 {}{}
}%
\makeatother
\begin{document}

\preprint{AIP/123-QED}

\title[High-Resolution Capacitance Dilatometry of Microscopically Thin Samples Using a Miniature Dilatometer.]{High-Resolution Capacitance Dilatometry of Microscopically Thin Samples Using  a Miniature Dilatometer}.

\author{R. Küchler}
\thanks{These two authors contributed equally.}
\email{kuechler@cpfs.mpg.de}
\affiliation{Max Planck Institute for Chemical Physics of Solids, Nöthnitzer Stra{\ss}e 40, 01187 Dresden, Germany}

\author{S. Panja}
\thanks{These two authors contributed equally.}
\affiliation{University of Augsburg, Universitätsstra{\ss}e 2, 86135 Augsburg, Germany}

\author{S. Wirth}
\affiliation{Max Planck Institute for Chemical Physics of Solids, Nöthnitzer Stra{\ss}e 40, 01187 Dresden, Germany}

\author{P. Gegenwart}
\affiliation{University of Augsburg, Universitätsstra{\ss}e 2, 86135 Augsburg, Germany}


\begin{abstract}
We present a novel application of our high-resolution capacitance dilatometer, specifically engineered for the precise characterization of quantum materials. These materials, which often appear as ultrathin, platelet-shaped crystals, are known for exotic phenomena such as superconductivity, topological order and quantum spin liquid. However, these crystals seldom reach macroscopic dimensions, making them unsuitable for conventional dilatometry techniques. By introducing a modified sample-mounting configuration, our design enables high-resolution measurements of thermal expansion and magnetostriction along in-plane crystallographic directions in samples with thicknesses well below 500 $\mu$m. Validation measurements using a Quantum Design PPMS system confirm reliable performance for a 300 $\mu$m-thick silver platelet, relatively hard ferromagnetic EuB$_6$ single crystals down to 50 $\mu$m, and a 40 $\mu$m-thin, soft AgCrS$_2$ single crystal. This advancement significantly broadens the applicability of capacitance dilatometry, providing a powerful platform for investigating emergent phenomena in reduced-dimensional quantum systems.
\end{abstract}

\maketitle

\section{\label{chap1}Introduction}


Capacitive dilatometry has emerged as a key technique in the study of thermodynamic properties of single-crystal solids. Its outstanding sensitivity allows for the precise detection of structural distortions associated with phase transitions such as superconductivity, magnetism, or charge ordering. Single-crystal solids often exhibit anisotropic physical properties. Capacitive dilatometry measurements can be conducted precisely along crystallographic axes, enabling accurate measurements along specific directions\cite{Steppke2013,Hafner2022,Debicki2024}. This directional resolution is essential for understanding anisotropic electronic or magnetic behavior. Furthermore, capacitance dilatometers operate reliably over a broad range of temperatures and can be used in high magnetic fields\cite{Kuechler2017,Kuechler2023}, making them ideal for probing materials under extreme conditions where novel quantum phases frequently emerge.

For all widely recognized capacitive dilatometers in use worldwide, the approach to sample installation remains fundamentally the same \cite{White1961, Griessen1973, Pott1983, Steinitz11986, Rotter1998, Schmiedeshoff2006, Neumeier2008, Park2009, Kuechler2012, Kuechler2016, Kuechler2017, Neumeier2022, Kuechler2023}. The sample is placed against the movable part of the dilatometer and then held in place with a mechanical clamp. This method requires compact, mechanically stable samples, typically in the millimeter range, that can stand freely and maintain their orientation during mounting. Consequently, very thin or fragile samples cannot be installed reliably and are generally unsuitable for use with these instruments.

This limitation is especially relevant for modern quantum materials. Many systems exhibiting unconventional superconductivity, exotic magnetism, topological phases, or quantum-spin-liquid behavior are available only as thin, platelet-shaped single crystals, often with thicknesses of 100 $\mu$m or less. Prominent material families, including cuprates, nickelates, and ruthenates, possess layered crystal structures with strong in-plane bonding and weak interlayer coupling\cite{Basov2017,Samarth2017,Zhang2017,Giustino2021,Berry2024}. Owing to their anisotropy and growth constraints, these materials rarely form compact crystals and are instead obtained as small platelets or thin films\cite{Basov2017,Pistawala2022}.

To date, precise capacitive dilatometry of thin platelets has been limited by the lack of a standardized mounting technique, a limitation we address in this work by introducing a novel mounting approach.

Thinner samples, clamped along their short axis, produce smaller length changes and weaker capacitance signals, reducing measurement sensitivity. Accurately measuring thermal expansion in micrometer-thick samples remains challenging, even with modern capacitive dilatometers capable of sub-Ångström resolution\cite{Schmiedeshoff2006, Neumeier2008, Park2009, Kuechler2012, Kuechler2016, Kuechler2017, Neumeier2022, Kuechler2023}. For example, a 10 $\mu$m-thick sample with relatively large $\alpha$ = $10^{-4} K^{-1}$ exhibits a length change of only $10$~\AA{} over 1 K, limiting the precision of absolute measurements.

\begin{figure}
\includegraphics[width=1.1\linewidth]{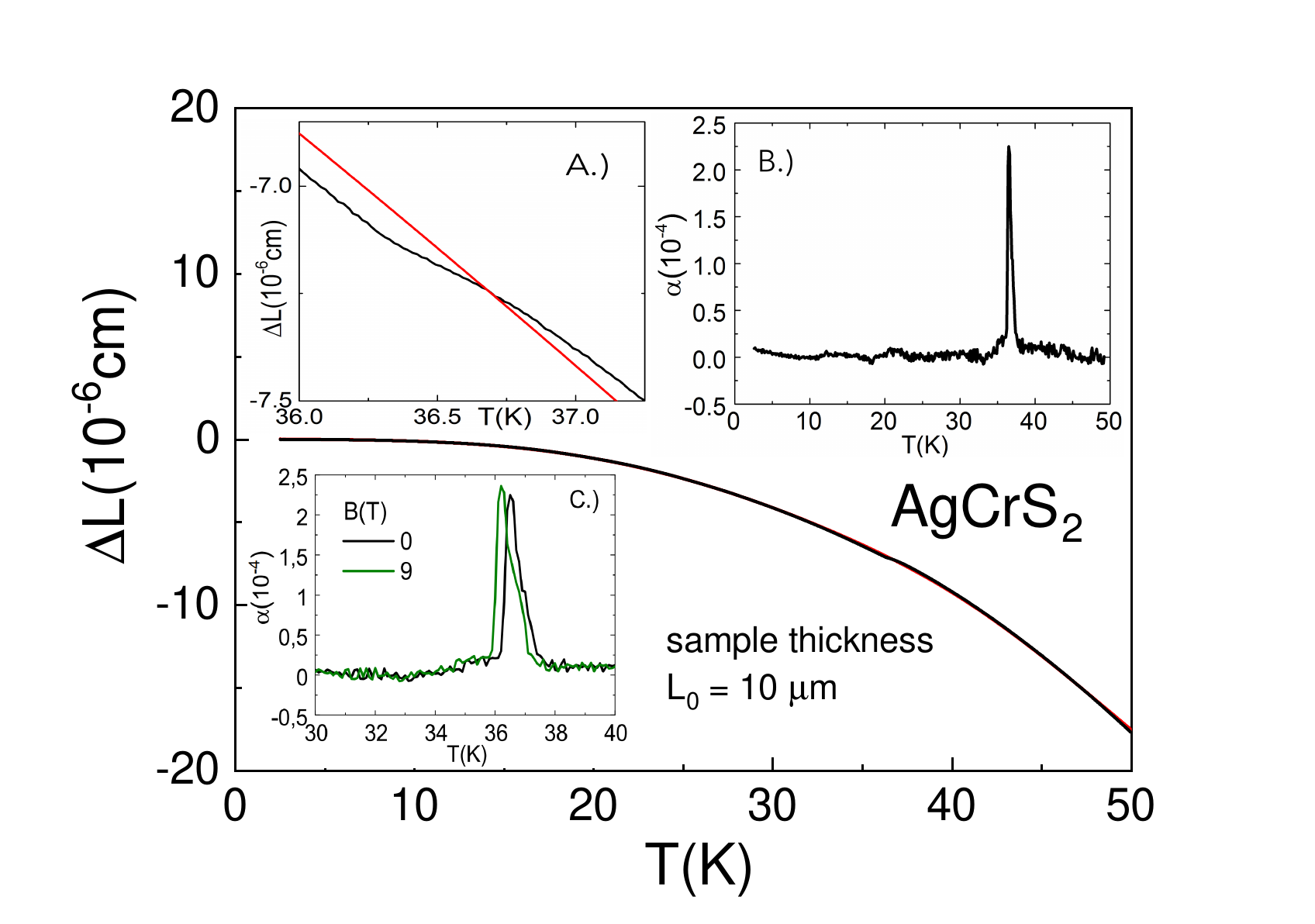}
 \caption{Main panel displays the measured length change $\Delta L(T)$ of the mounted 10 $\mu$m-thick AgCrS$_2$ sample (black curve), along with the background signal from the empty cell (red curve) for comparison. Inset A) shows an expanded view of both curves in between 36 K $\leqq T \leqq$ 37.25 K, illustrating detailed changes near the phase transition. Inset B) presents $\alpha(T)$ of AgCrS$_2$, obtained after subtraction of the background signal from the empty cell. Note that the pronounced background noise for the very thin sample ($L_0$ = 10 $\mu$m) results essentially from the $1/L_0$ scaling of $\alpha(T) = (1/L_0) dL/dT$, which amplifies the effect of any length fluctuations. For such a small absolute expansion, these instrumental noise sources appear much larger in $\alpha(T)$ than for thicker samples. Inset C) shows a comparison of $\alpha(T)$ measured in zero magnetic field and under an applied magnetic field of 9 T.}
 \label{Fig0}
\end{figure}

We recently performed thermal expansion measurements on a 10 $\mu$m-thick single crystal of AgCrS$_2$ using our miniaturized dilatometer \cite{Kuechler2023} integrated into the Physical Property Measurement System (PPMS). At the time of the experiment, thicker single crystals were not available. AgCrS$_2$ is a layered material known to undergo a structural phase transition near 37~K, that is closely associated with the onset of magnetic order \cite{Damay2011}. This transition involves a lowering of the symmetry towards a ferroelectric monoclinic phase \cite{Damay2013,Meher2014} indicative of strong spin-lattice coupling and cooperative ordering phenomena. Measurements were carried out perpendicular to the layers, along the thin (10~$\mu$m) sample axis. Note that this is different from the measurements on thin EuB$_6$ and AgCrS$_2$ samples reported below where thermal expansion was measured perpendicular to the thin sample dimension.

Fig.\ \ref{Fig0} shows the raw length-change data of the AgCrS$_2$ sample alongside the background signal of the empty measurement cell. At first glance, the two curves appear nearly identical. However, a step-like anomaly associated with the structural phase transition becomes clearly visible between 36 K and 37 K (Inset A). After subtraction of the cell background, the intrinsic sample response becomes apparent, revealing a step height of $\Delta L =$ 15~\AA{} at the transition. The thermal expansion coefficient $\alpha(T)$ was determined over the full temperature range of 2–50 K using interval differentiation with a 0.1 K step (Inset B). Outside the phase transition region, $\alpha(T)$ remains essentially zero, reflecting the limited resolution of our dilatometer for such an extremely small sample of only 10 $\mu$m thickness. The absolute length changes of the sample, together with the background contribution of the dilatometer cell, approach the magnitude of the background itself, making precise determination of $\alpha(T)$ unreliable. Despite this limitation, the structural phase transition is clearly observed as a sharp $\lambda$-type anomaly at 36.5 K. Under an external magnetic field, the transition shifts to lower temperature, in agreement with previous reports\cite{Apostolova21,Kanematsu21}. In particular, a field of 9 T reduces the transition temperature by $\approx$ 0.3 K (Inset C). These results demonstrate that ultra-thin samples are sufficient for qualitative detection of structural anomalies, but accurate quantitative measurements of $\alpha(T)$ require larger single crystals, for which background subtraction is  more reliable.

An outstanding, yet often desirable, task is to measure thermal expansion of ultra-thin samples along its in-plane orientation. Assuming isotropic sample properties, measuring the sample along
its longer dimension may help resolving the thermal expansion signal considerably. For anisotropic sample properties as expected for layered materials, such a measurement may provide valuable additional insight into the material's properties. However, this configuration introduces new and significant challenges, including precise alignment, mechanical fragility, and increased sensitivity to mounting-induced artifacts, all of which must be carefully addressed to ensure reliable measurements.

Thin samples of only a few tenths of $\mu$m thickness are often flexible and fragile and are difficult to mount without introducing mechanical stress. They can flex or bow under even the light dilatometer spring force. Any bending or warping under mounting pressure causes measurement artifacts.
As described in detail in Chapter II.B of Ref.\cite{Kuechler2023}, our dilatometer design with two 0.25-mm-thick leaf springs, applies a modest spring force of approximately 3–4 N to the sample. For millimeter-sized samples, such a weak load mostly does not affect the intrinsic material properties. However, for ultra-thin samples with thicknesses of about 50 $\mu$m, the resulting uniaxial pressure becomes significantly larger and cannot be neglected. These pressure-induced effects must therefore be considered in the data interpretation. A detailed discussion of the resulting pressure-dependent contributions is provided in Chapter III.B for the specific case of the measured 50-$\mu$m-thick EuB$_6$ single crystal.

In the 2017 Editor’s Pick, we introduced “The world’s smallest high-resolution capacitive dilatometer.” Although capacitive cells of comparable dimensions have been reported, our design achieved an unprecedented resolution for a device of this scale, capable of detecting relative length changes as small as $0.01$~\AA \cite{Kuechler2017, Kuechler2023}. The extreme miniaturization of our dilatometer has enabled its integration into several new experimental platforms, including \textit{in-situ} sample rotation within Quantum Design’s Physical Property Measurement System (PPMS) and their dilution refrigerator insert\cite{Kuechler2017, Kuechler2023}. Our unique dilatometer design also allows for the application of additional uniaxial pressure by replacing the so-called “body” component. This interchangeable-body concept enables the mini-dilatometer to perform high-resolution measurements of thermal expansion and magnetostriction, even with the application of substantial uni-axial stress\cite{Kuechler2023}.

In this publication, we demonstrate that a modification of our sample-mounting configuration enables measurements of microscopically thin, platelet-shaped samples along their larger dimension. This advancement significantly extends the applicability of capacitance dilatometry, allowing for the investigation of a wider range of modern quantum materials as well as their anisotropic properties.

\section{\label{chap2}New Application: Measuring microscopically thin samples
}
 \subsection{\label{chap2a} The commonly used Mounting Procedure}

\begin{figure}
\includegraphics[width=0.99\linewidth]{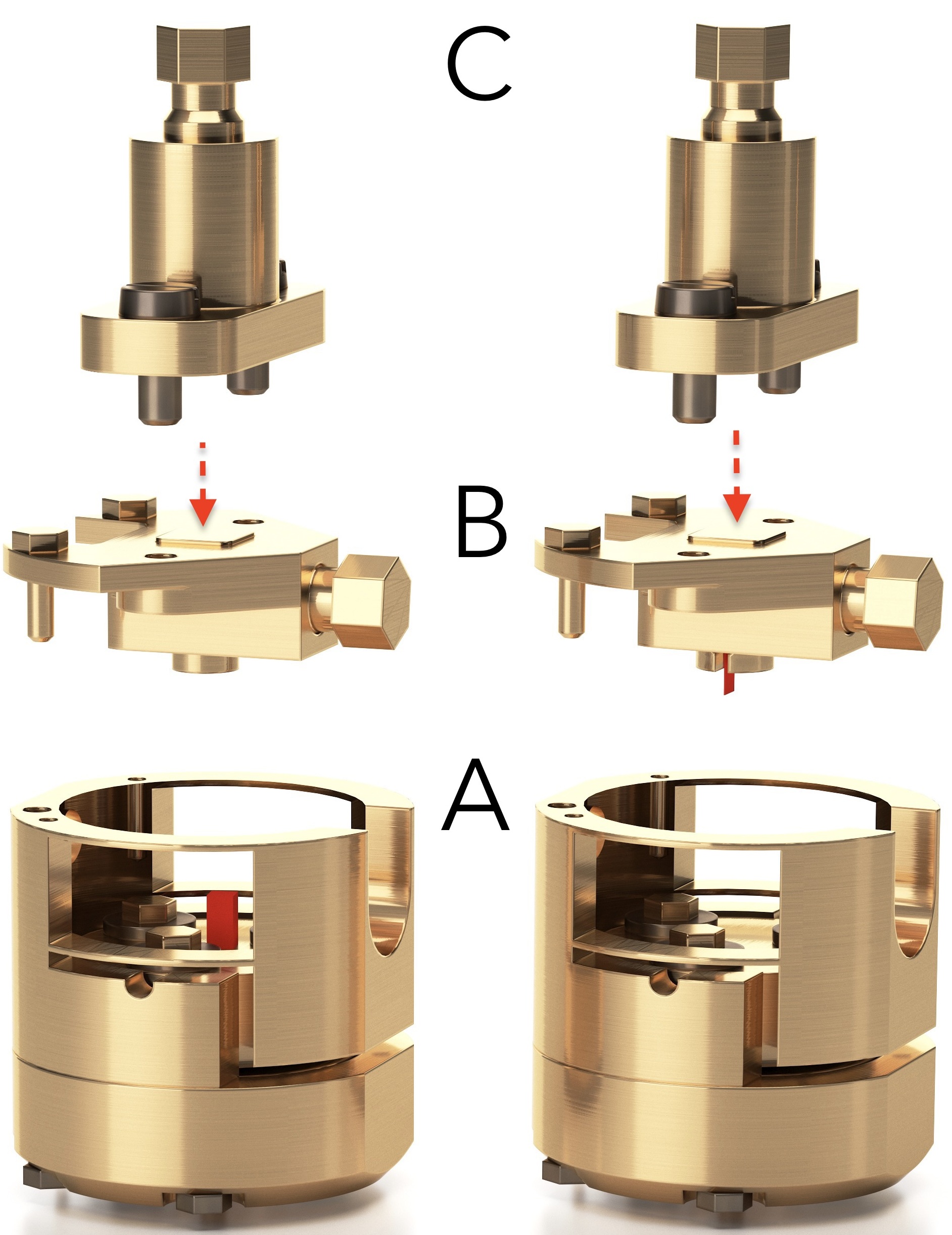}
 \caption{Left: Common sample mounting procedure with cube-shaped stamp (indicated by a red
 arrow).  Dilatometer consists of three main parts: (A) main body with sample (dark red)
 positioned in its center, (B) cover, and (C) sample-adjusting tool. Right: Sample mounting
 procedure using a slotted cube-shaped stamp, shown here with a mounted sample of 50 $\mu$m
 thickness (dark red).}
 \label{Fig1}
\end{figure}

The image in Fig.\ \ref{Fig1}, left, illustrates the standard procedure for inserting a mm-sized sample into the dilatometer. The sample (red cuboid part) is inserted vertically into the center of the body from above. To ensure proper positioning, the sample must have a sufficiently large cross-sectional area to remain upright and must be aligned along the vertical measurement axis z. Once the sample is in place, the cover (B) and adjustment tool (C) are attached. This method is generally effective for samples with a cross-section no smaller than $0.4 \times 0.4$~mm$^{2}$. For samples with smaller cross-section, however, maintaining a stable upright position during installation becomes increasingly difficult, as they are prone to tipping or shifting before clamping can be completed. Clamping is achieved by gently tightening the adjustment screw (located on the top of sample-adjusting tool C), which does not act directly on the sample. Instead, it applies force to a cube-shaped stamp (see red arrows in Fig.\ \ref{Fig1}), which is constrained to move only vertically within the cover (B). This design prevents any rotational or angular displacement of the sample during clamping, preserving its crystallographic orientation. Once the sample is clamped, the locking screw on the right side of part (B) secures the cubic stamp in place, allowing the sample-adjusting tool (C) to be removed.

Mounting ultra-thin samples can become a significant challenge. To address this, several alternative mounting strategies have been explored, including the stacking of multiple thin crystals and the attachment of samples to substrates with well-characterized thermal expansion behavior. However, these approaches introduce potential measurement errors, as adhesive materials can affect the thermal and mechanical coupling between the sample and its surroundings, thereby reducing the reliability of the data. Unless crystals within a stack are perfectly bonded or compressed, microscopic air or vacuum gaps may exists between them and give rise to inhomogeneous heat transfers and temperature gradients between the samples. Furthermore, internal friction or slippage between samples can generate artifacts that do not accurately reflect the intrinsic thermal expansion of the material. As a result, the measurement quality may strongly deteriorate.

\subsection{\label{chap2b} Evaluation of a Side-Support Mounting Method for Platelet-Like Crystals in a Capacitance Dilatometer}

An alternative sample mounting approach was tested in which a thin, platelet-like crystal was positioned vertically by sandwiching it between two rectangular support blocks. These blocks were placed laterally on either side of the crystal to hold it upright during the critical moment when the cubic adjustment stamp of the dilatometer clamps the sample. Copper blocks were fabricated in appropriate dimensions for this purpose, with their heights chosen to be approximately 0.5 to 0.7 times the length of the crystal. This ratio was intended to provide adequate stability while avoiding interference with the dilatometer’s mounting mechanism. To make it possible to remove the support blocks after aligning the crystal, thin threads were attached to them. However, a significant challenge with this method was the tendency of the support blocks to shift during the application of clamping force, often leading to slippage or tilting of the crystal. Despite careful handling, this instability limited the effectiveness and reproducibility of the approach.

\subsection{\label{chap2c}Improved Sample Mounting Technique Using a Slot-Based Stamp for Platelet-Shaped Crystals}

\begin{figure}
\includegraphics[width=0.90\linewidth]{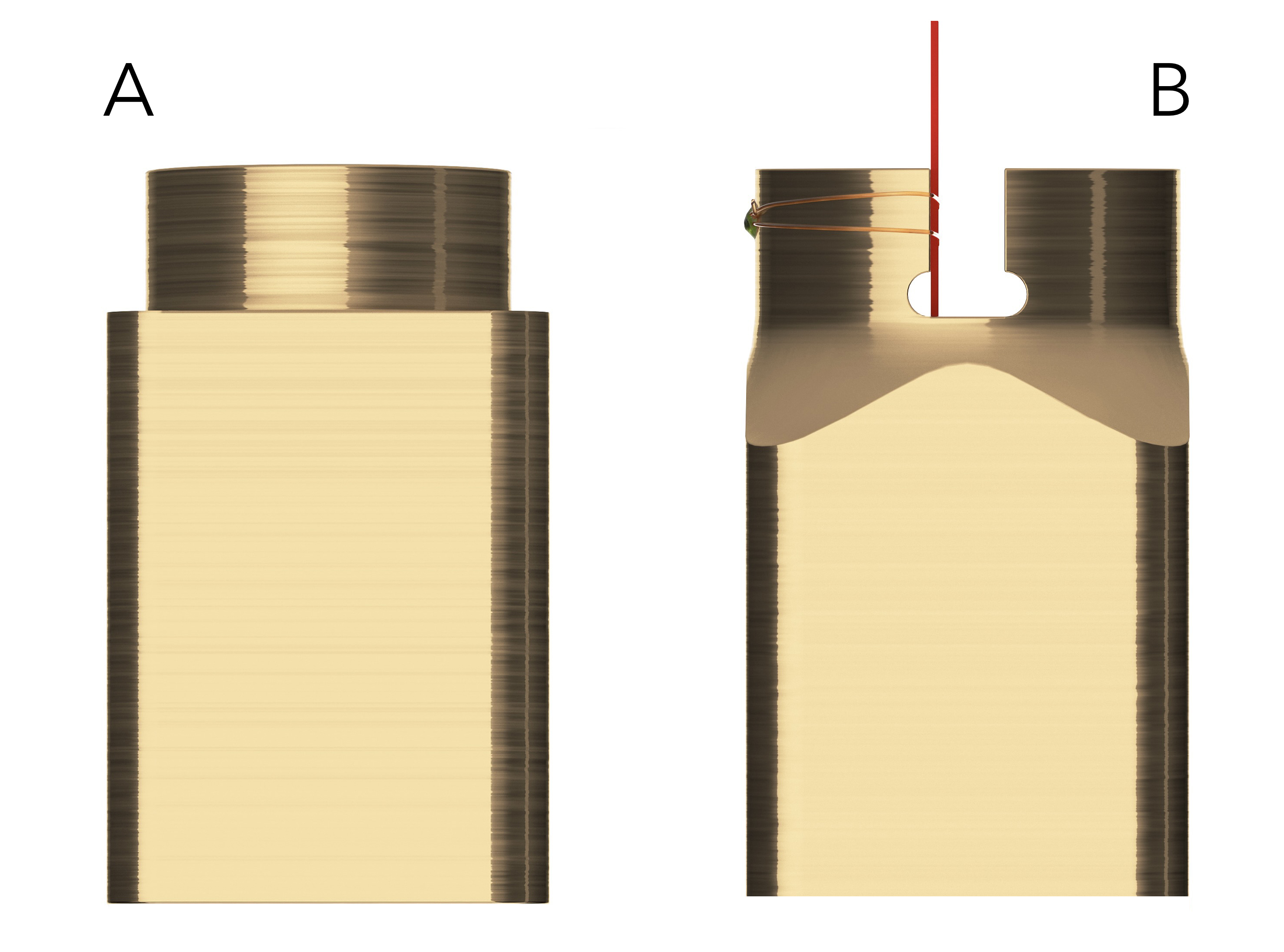}
 \caption{A) Standard cube-shaped mounting stamp commonly used for sample mounting. B) Slotted cube-shaped stamp with an ultra-thin sample (red) mounted. Two 30~$\mu$m-thick PTFE-coated silicon threads are looped around the sample.}
 \label{Fig2}
\end{figure}

We introduce an optimized method for mounting thin, platelet-shaped single crystals in our dilatometer, designed to improve alignment precision and mechanical stability. The core of this approach is a slotted mounting stamp, fabricated with a narrow 0.5 mm wide slot produced by precision wire erosion. During fabrication, the wire is rotated by $90\degree$ at its maximum penetration depth to create a C-shaped undercut at the lower end of the slot, as illustrated in Fig.\ \ref{Fig2}B. This geometry ensures precise, full-length contact between the sample and the slot side-wall, enhancing alignment accuracy and mechanical stability. To secure the sample within the slot, a 30~$\mu$m-thick PTFE-coated silicon thread is looped around the sample and fixed to the outer side of the stamp using adhesive (see Fig.\ \ref{Fig2}B). For samples exceeding 1 mm in length, two such loops are used to further enhance mechanical stability during stamp handling and installation (Fig.\ \ref{Fig2}B).

Once secured, the stamp containing the mounted sample is inserted vertically through the cubic opening of the cover piece (B) and held in place using the right-side lock screw of cover piece (B), as shown in Fig.\ \ref{Fig1}B. After initial positioning, the cover (B) is reattached to the dilatometer body (A). The lock screw is then loosened, and the stamp is carefully lowered by hand until the sample gently contacts the bottom surface of the dilatometer body (A), using controlled force to avoid damaging the sample. With the sample in position, the standard mounting procedure can be continued as described above.

The sample mounting can be inspected using a microscope to verify that the sample remains straight and has not bent under the applied spring force of typically 4~N\cite{Kuechler2023}. This slot-based method significantly reduces the risk of sample tipping or misalignment. As will be shown below, it provides a robust and user-friendly solution for routine measurements on ultra-thin samples.

\section{\label{chap3}Test measurements on thin samples}

In the following we present thermal expansion measurements on thin samples using the In-situ dilatometer probe described in detail in Ref.\cite{Kuechler2023}. Because of its small size, this dilatometer can be rotated inside the PPMS around a horizontal axis. For the test measurements reported here, experiments were carried out in zero magnetic field. Application of a magnetic field is not expected to pose any problem as long as the magnetic moment of the sample is small enough to not exert excessive mechanical forces on the sample.

\subsection{\label{chap3a} Test measurement on 300-micron thin silver platelet}

\begin{figure}
\includegraphics[width=0.98\linewidth]{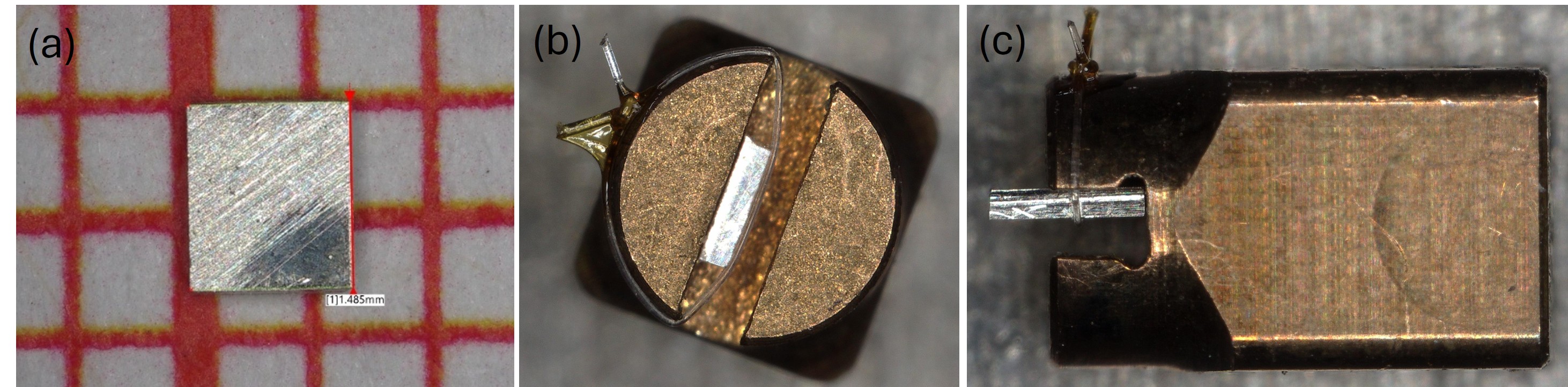}
 \caption{(a) A 300 $\mu$m thin rectangular silver plate mounted within the narrow slot of our custom-made cubic mounting stamp. Figures (b) and (c) show the top and side views of the mounted silver platelet, respectively.}
 \label{Fig5}
 \end{figure}

A high-purity silver sheet (99.95\%, Goodfellow GmbH) with a thickness of 300 $\mu$m was cut into a rectangular plate measuring 1.48~mm in length, and polished to ensure parallel edges, as shown in Fig.\ \ref{Fig5}(a). High-purity silver has a Vickers hardness of roughly 0.24 GPa\cite{Staff2013Silver}, which is very soft compared to most engineering metals. The thin silver plate was then securely mounted within the narrow slot of the mounting stamp. To achieve proper parallel alignment, the plate was gently leaned against one wall of the slit and carefully tied in place using a pre-positioned loop of 30 $\mu$m-thick PTFE-coated silicone wire. For additional mechanical stability, the wire was secured with adhesive to the rear side of the mounting stamp. Top and side views of the sample mounted within the stamp are shown in Figs.\ \ref{Fig5}(b) and \ref{Fig5}(c), respectively.

To accurately determine the intrinsic length change of the sample, the contribution from the dilatometer cell must be subtracted. This was achieved by measuring a high-purity copper
(Cu) reference sample that matches the length of the thin silver plate. The relative length change $\Delta L(T)/L_0$ and the linear thermal expansion coefficient $\alpha$(T) of the thin silver plate were then determined by subtracting the so-determined cell contribution, following the procedure outlined in Ref.\cite{Kuechler2012}.

\begin{figure}
\includegraphics[width=0.98\linewidth]{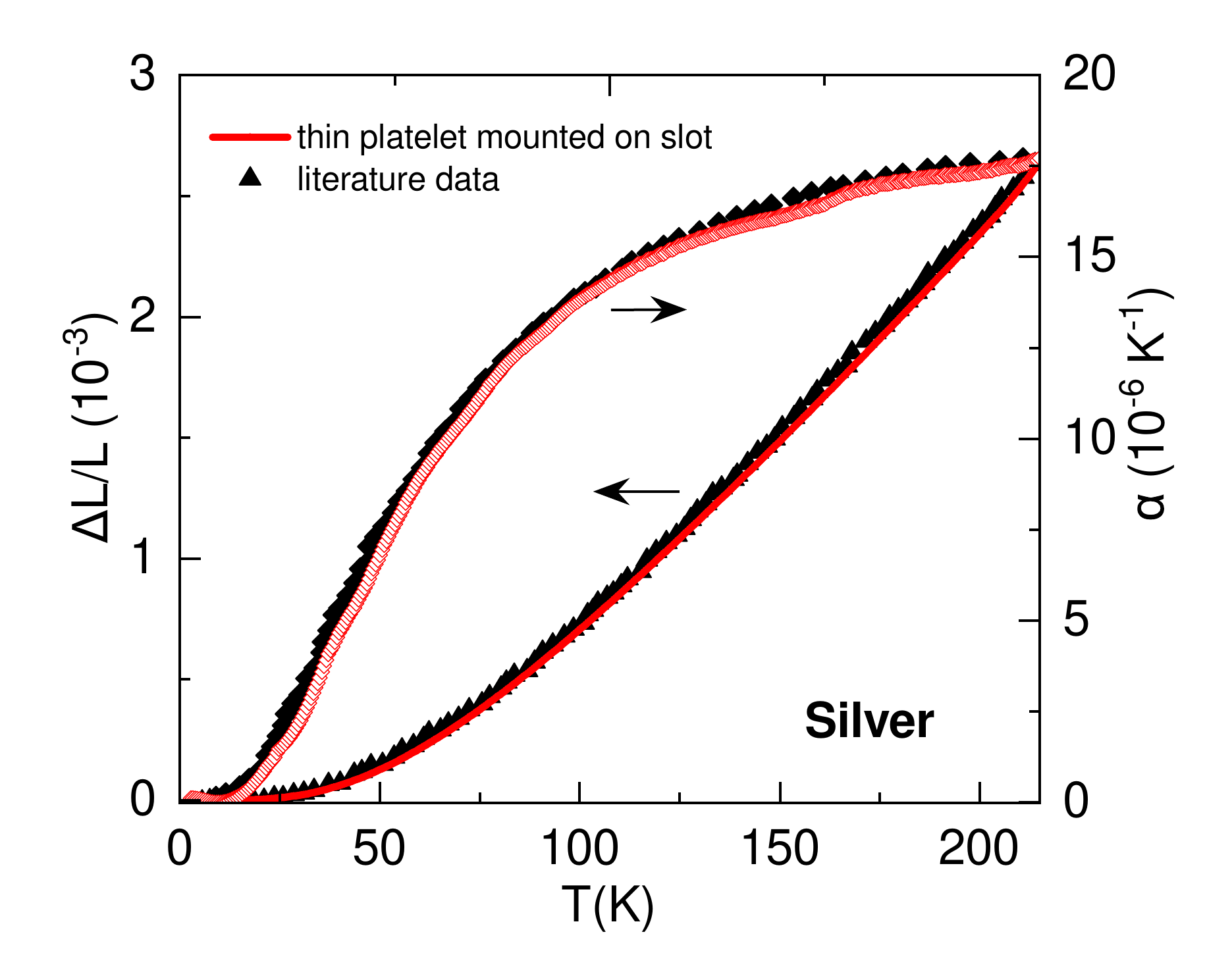}
 \caption{Comparison of the relative length change, $\Delta L(T)/L_0$, and the thermal expansion coefficient, $\alpha(T)$, as a function of temperature for the 300 $\mu$m silver plate measured while mounted in the narrow slot, alongside literature data from Ref.\cite{Kuechler2012}. Very good agreement is observed across the entire temperature range, confirming the reliability of the measurement setup.}
 \label{Fig6}
\end{figure}

Fig.\ \ref{Fig6} presents the resulting $\Delta L(T)/L_0$ and $\alpha(T)$ of the silver plate
as function of temperature. The results show very good agreement with previously reported literature data for bulk silver (Ref.\cite{Kuechler2012}). This consistency confirms the effectiveness of our slot-based sample mounting approach, as demonstrated by measurements on a well-characterized reference metal.

\subsection{\label{chap3b} Test measurements on ferromagnetic semimetal EuB$_6$}

\begin{figure}[b]
\includegraphics[width=0.98\linewidth]{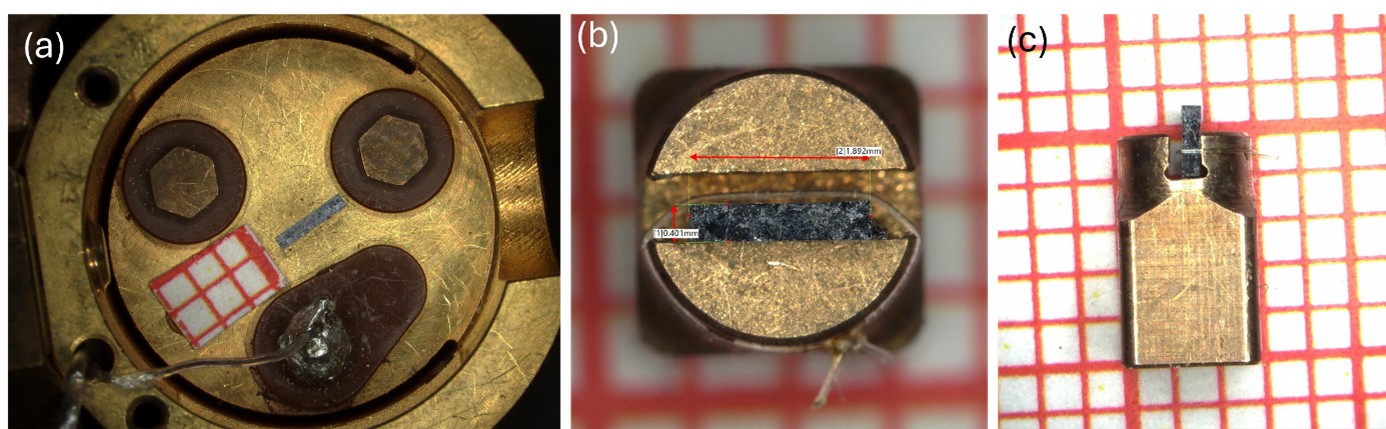}
 \caption{(a) EuB$_6$ crystal (length $\approx$1.60~mm, thickness $\approx$ 400~$\mu$m) positioned upright on the bottom plate of the dilatometer, mounted as described in section \ref{chap2a} (standard mounting procedure). (b) Top view of the same 400 $\mu$m-thick EuB$_6$ crystal mounted within the stamp slot. (c) Side view of the mounted crystal.}
 \label{Fig7}
\end{figure}

To further validate the reliability of the proposed slot-based mounting technique, thermal expansion measurements were conducted on thin plates and ultra-thin platelets prepared from EuB$_6$ single crystals. We emphasize that EuB$_6$ crystallizes in a body-centered cubic structure and hence, uniform properties are expected for measurements along three main crystallographic axes. Upon cooling, it undergoes two successive transitions: The first one at $T_1$ = 15.4~K is related to the percolation of magnetic polarons while the second one at $T_2$ = 12.6~K indicates the onset of ferromagnetic order in EuB$_6$ \cite{ZFisk1997,Suellow2000}. The charge delocalization near the percolation transition as well as the spontaneous magnetization resulting from local moment exchange gives rise to significant contributions to the thermal expansion \cite{Manna2014}.We note that EuB$_6$ is a hard, stiff, and relatively strong material (Vickers hardness $\approx$ 28.5 - 28.9 GPa, Young’s modulus $\approx$ 225 - 240 GPa) \cite{Futamoto1979,Cahill2019,Karre2022} and therefore well suited for our measurements.

\begin{figure}
\includegraphics[width=0.98\linewidth]{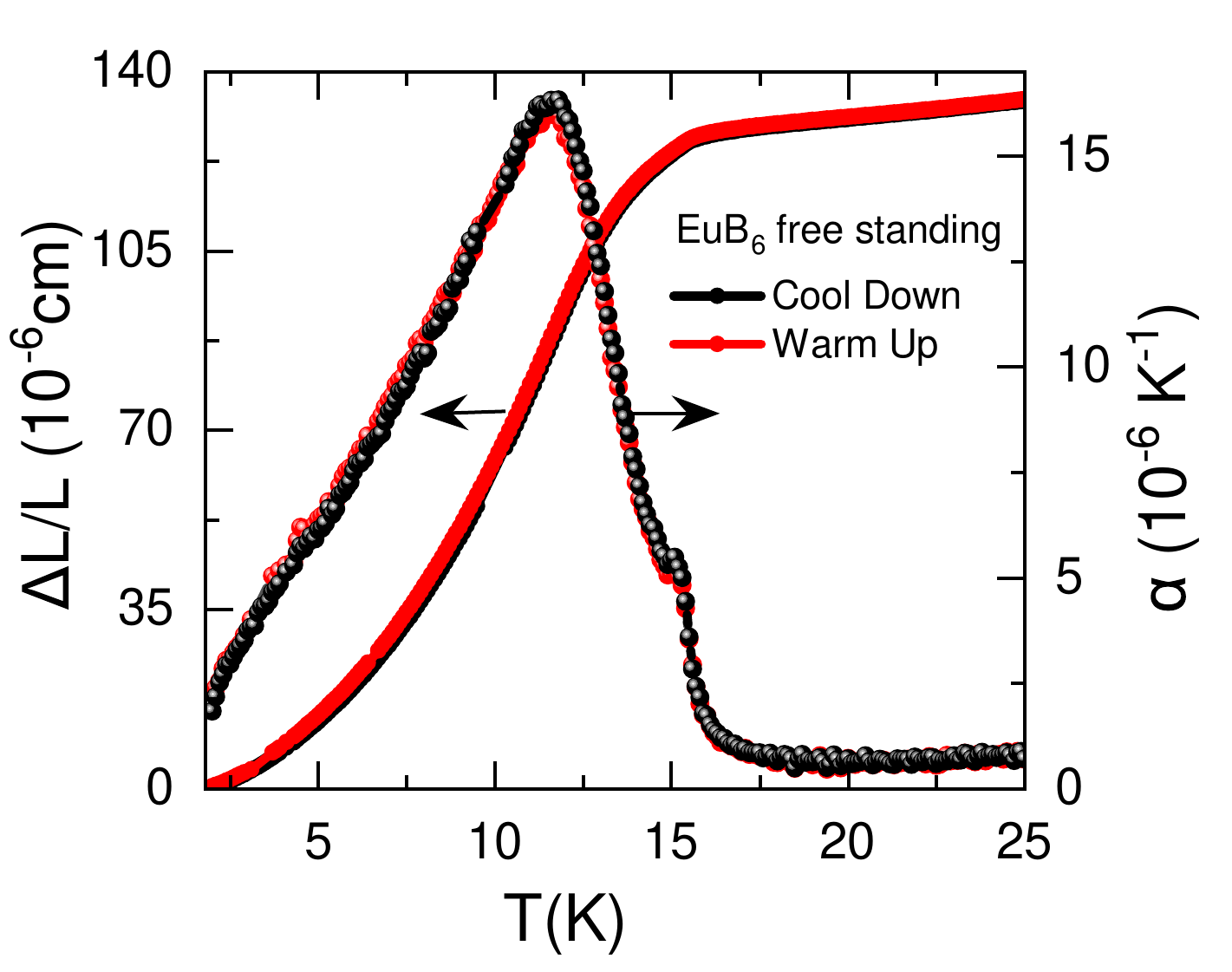}
 \caption{Temperature-dependent relative length change $\Delta L(T)/L$  and thermal expansion coefficient $\alpha(T)$ for a free-standing 400~$\mu$m EuB$_6$ sample.}
 \label{Fig8}
\end{figure}

The initial measurement was performed using the conventional mounting method, in which a 400 $\mu$m-thick EuB$_6$ single crystal (length $\approx$ 1.60~mm) was positioned as a free-standing rectangular piece, as shown in Fig.\ \ref{Fig7}(a). Fig.\ \ref{Fig8} presents the relative length changes $\Delta L(T)/L_0$ and the corresponding thermal expansion coefficient $\alpha(T)$ measured during both heating and cooling cycles.

\begin{figure}[b]
\includegraphics[width=0.98\linewidth]{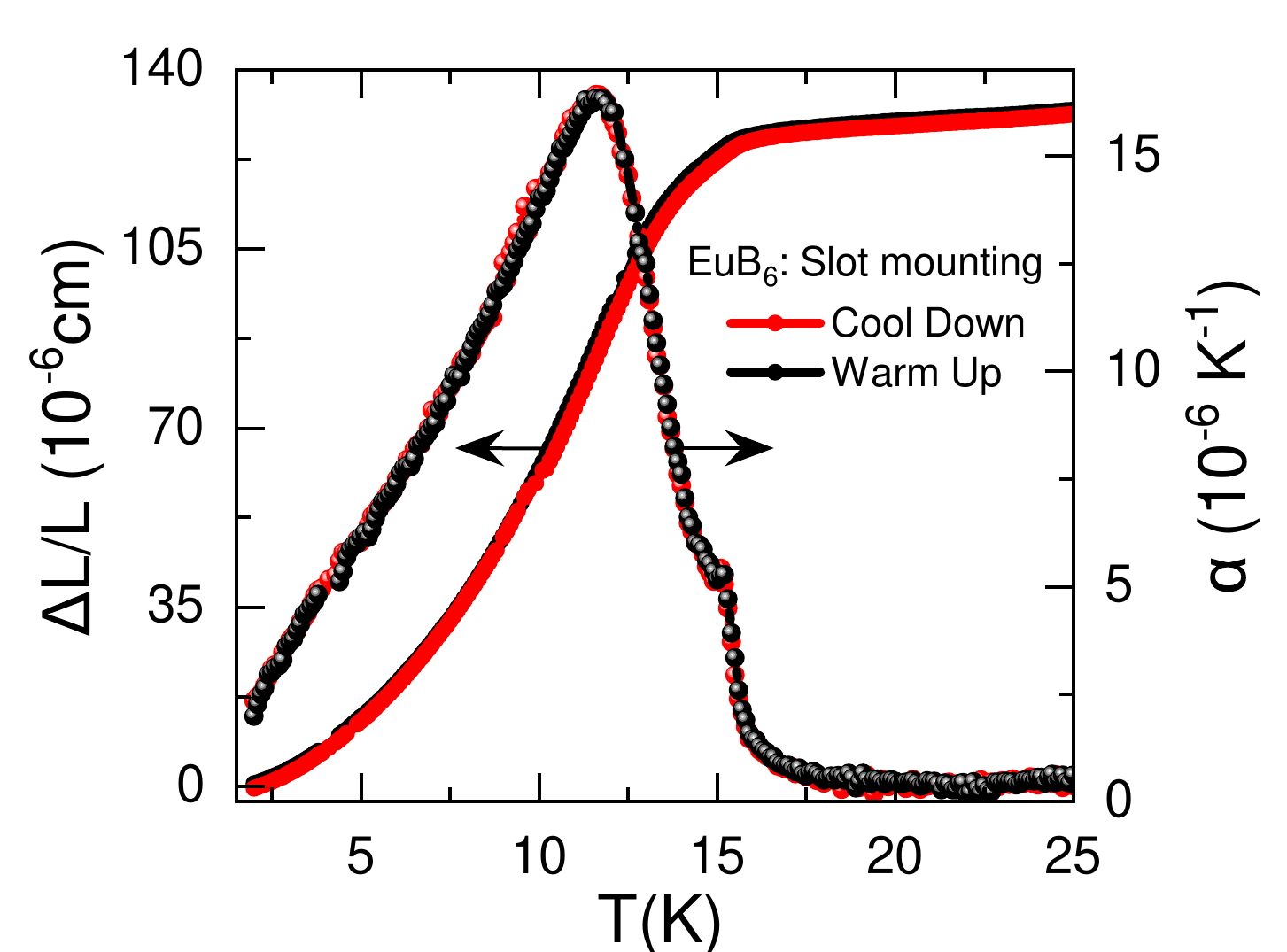}
 \caption{Measurement of relative length change $\Delta L/L_0$ and thermal expansion coefficient $\alpha(T)$ for a 400~$\mu$m EuB$_6$ single crystal mounted within the narrow slot of the cubic mounting stamp.}
 \label{Fig9}
\end{figure}

\begin{figure}
\includegraphics[width=0.98\linewidth]{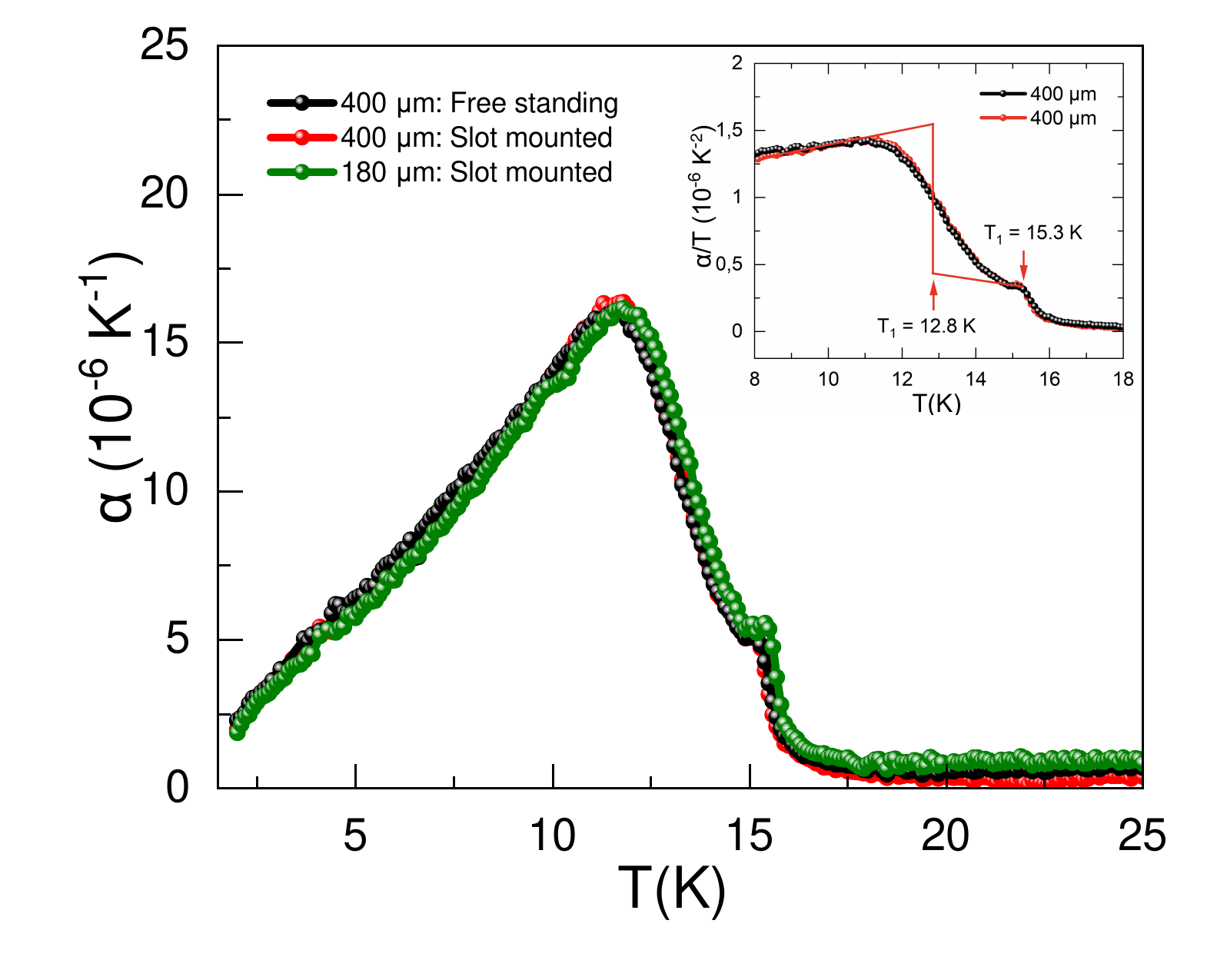}
\caption{Comparison of the thermal expansion coefficient $\alpha(T)$ from three independent measurements: An identical 400~$\mu$m-thick EuB$_6$ single crystal in two configurations namely i) free-standing and ii) slot-mounted. iii) A 180~$\mu$m-thick EuB$_6$ sample from the same batch \#1 in slot-mounted configuration. Inset shows the same data for the 400~$\mu$m EuB$_6$ sample in a plot $\alpha(T)/T$ vs.\ $T$, where the two distinct anomalies at $T_1$ = 15.3 K and $T_2$ = 12.8 K were identified by a sharp transition and an entropy-conserving equal-area construction, respectively.}
 \label{Fig10}
\end{figure}

The same crystal was then mounted again using the proposed slot-based technique, see Figs.\ \ref{Fig7}(b) and (c), and remeasured under otherwise identical conditions. Results on the  slit-mounted sample are shown in Fig.\ \ref{Fig9}. Notably, both mounting techniques yield identical results (see Fig.\ \ref{Fig10}), demonstrating its reliability and reproducibility. In both cases, no thermal hysteresis is observed, indicating excellent thermal coupling between the sample and the dilatometer.

In agreement with previous thermal expansion studies on bulk samples \cite{Manna2014}, two distinct anomalies are observed in the lattice expansivity, Figs.\ \ref{Fig8}, \ref{Fig9} and \ref{Fig10}: A sharp transition occurs at $T_1$ = 15.3~K, followed by a pronounced maximum slightly below 12~K. An entropy-conserving, equal-area construction applied to the $\alpha(T)/T$ vs.\ $T$-data (see inset of Fig.\ \ref{Fig10}) yields a second transition temperature of $T_2$ = 12.8~K. It is important to note that the resolution of these two distinct anomalies requires samples of exceptionally high quality \cite{Manna2014}.

\begin{figure}[b]
\includegraphics[width=0.98\linewidth]{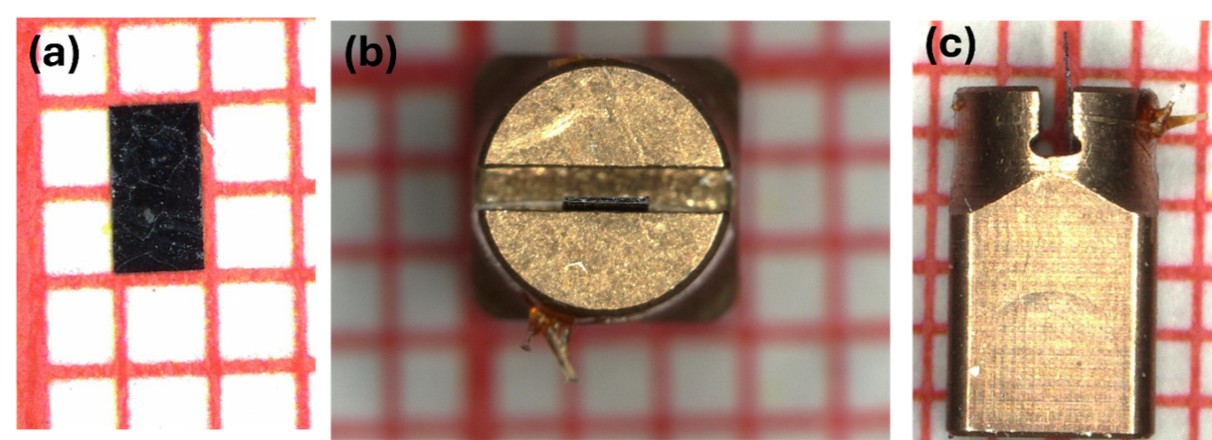}
 \caption{ (a) 50 $\mu$m EuB$_6$ ultra-thin single crystal with parallel edges. (b) Top view of
 the EuB$_6$ crystal mounted inside the narrow slot. (c) Side view of the mounted crystal.}
 \label{Fig11}
\end{figure}

To further test the performance of the slit-based mounting technique on even thinner samples, we measured a EuB$_6$ single crystal of thickness $\sim$180~$\mu$m, the thinnest sample available from the same batch \#1 as the previously measured 400~$\mu$m-thick crystal. Fig.\ \ref{Fig10} provides a comparison of the thermal expansion data of the 180~$\mu$m slit-mounted sample to the results of the 400~$\mu$m sample using both mounting techniques. All
measurements exhibit excellent agreement, including the very thin single crystal.

Encouraged by these results, we tested the applicability of the slot-mounting technique for an even thinner sample. A rectangular EuB$_6$ single crystal with a thickness of $\sim$50$\mu$m, obtained from a different batch than the previously studied samples, was carefully polished to ensure parallel edges, as shown in Fig.\ \ref{Fig11}(a). This ultra-thin platelet was subsequently mounted in the slot of the custom-designed stamp and secured using a single knot, as illustrated in Figs.\ \ref{Fig11}(b), (c).

\begin{figure}
\includegraphics[width=0.98\linewidth]{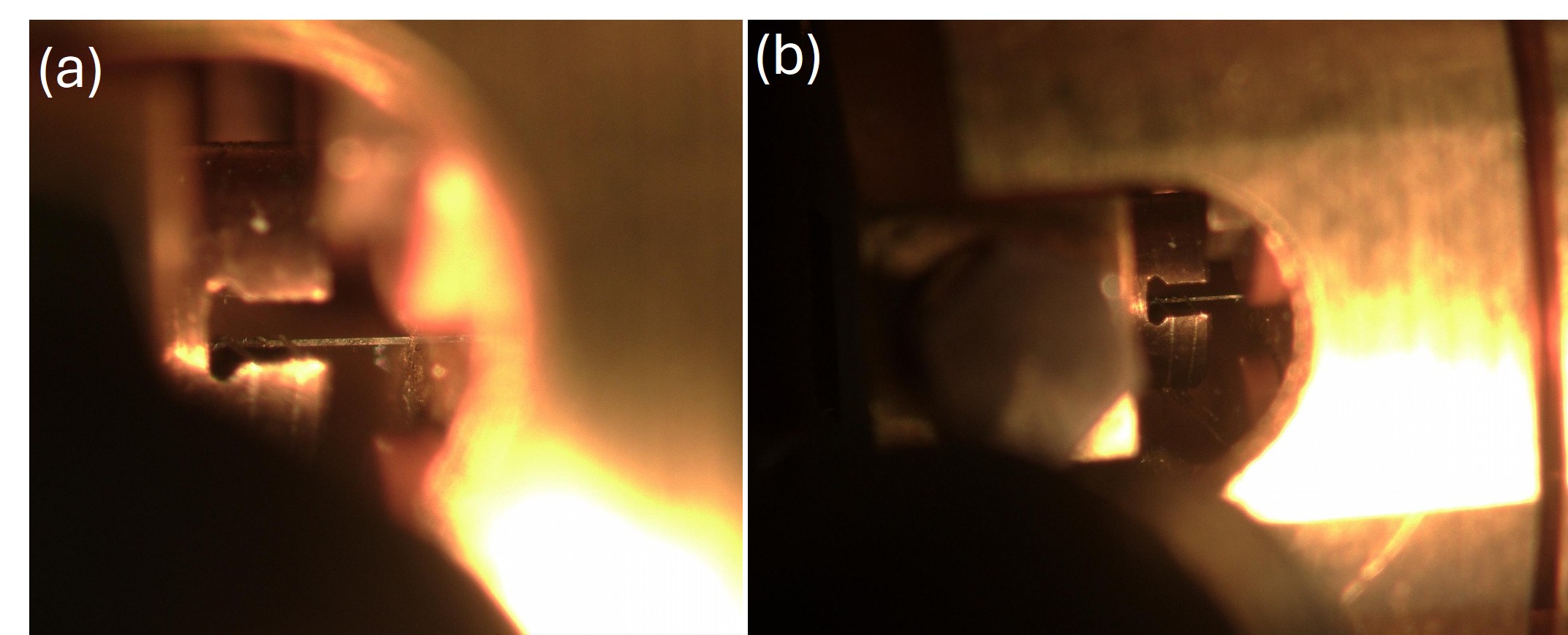}
 \caption{Optical microscope image of a 50~$\mu$m thin EuB$_6$ single crystal mounted in a miniature dilatometer. The crystal is positioned within the slot of the cubic stamp and secured using a 30~$\mu$m-thick PTFE-coated silicon thread. The initial capacitance is 20~pF, corresponding to an applied force of approximately 4~N \cite{Kuechler2023}. No bending or visible deformation of the crystal is observed under these conditions.}
 \label{Fig15}
\end{figure}

Accurate thermal expansion measurements using capacitive dilatometry rely on precise control over both the geometry and mechanical conditions of the sample. Prior to measurement, the sample was examined utilizing an optical microscope and verified to be straight and free of any bending (see Fig.\ \ref{Fig15}). This step is essential, as pre-existing bends can introduce uncertainty in the initial length and cause uneven deformation during thermal cycling, both of which can lead to significant artifacts in the measured signal. By confirming that the sample was flat and unbent, we ensured that the observed length changes reflect intrinsic thermal
expansion, rather than mechanical relaxation effects. This careful preparation enhances the reliability and interpretability of the dilatometric data, particularly for thin and delicate samples.

\begin{figure}
\includegraphics[width=0.98\linewidth]{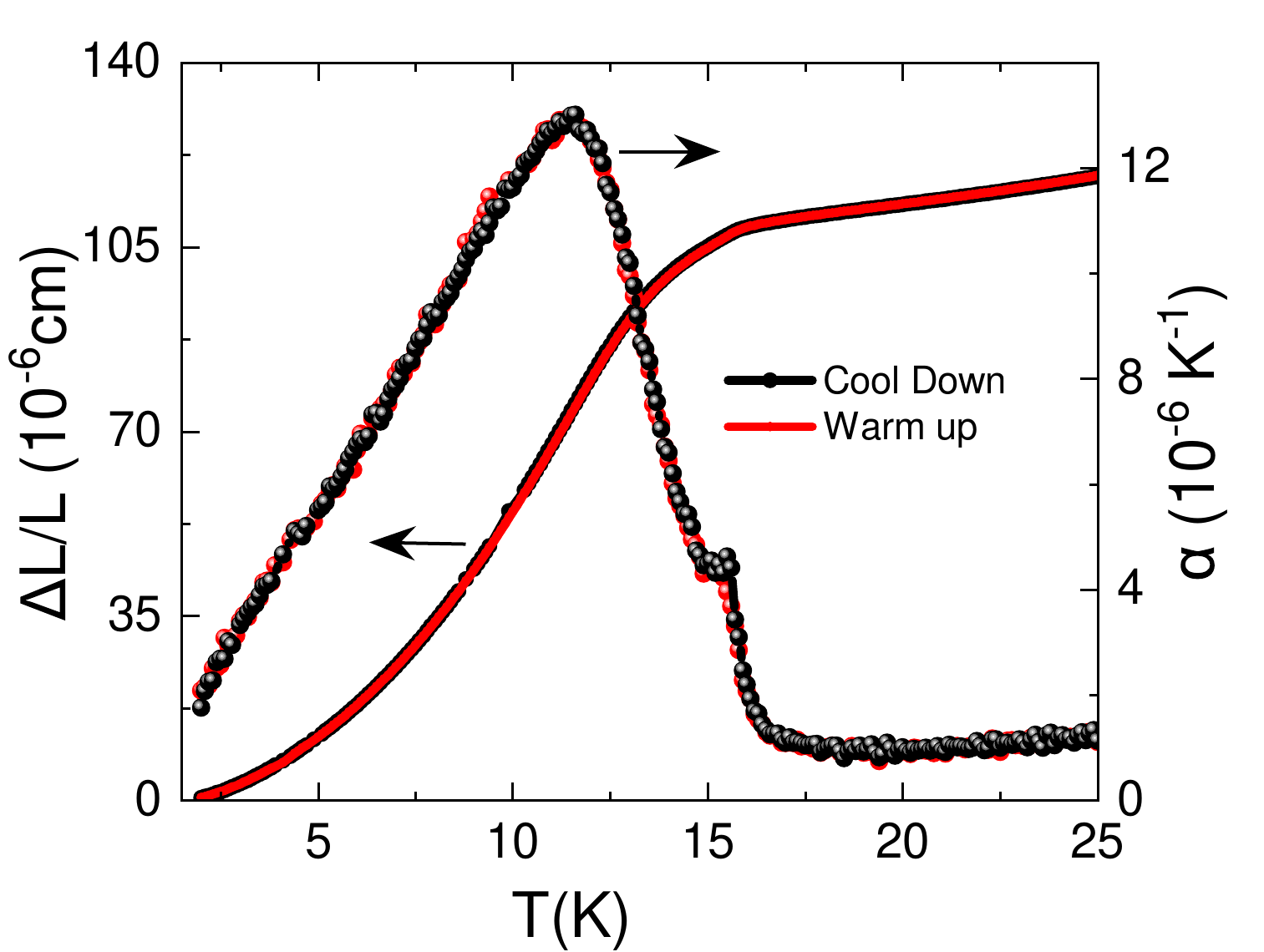}
 \caption{Relative length change $\Delta L(T)/L_0$ and  thermal expansion coefficient $\alpha(T)$ for a 50~$\mu$m EuB$_6$ thick single crystal (batch \#2) mounted inside the slot of the custom-made cubic mounting stamp.}
 \label{Fig12}
\end{figure}

$\Delta L(T)/L_0$ and $\alpha(T)$ measured during both the warm-up and cool-down cycles are presented in Fig.\ \ref{Fig12}. The shape of the measurement curves and the characteristics of the two observed anomalies are identical to those observed for the thicker samples. As before, no hysteresis is seen, indicating excellent thermal coupling of the sample. However, the measured values of $\Delta L(T)/L_0$ are only about 0.8 of the values obtained for thicker samples. This reduction may be attributed either to the fact that the 50~$\mu$m-thick sample originates from a different batch \#2, or to higher uniaxial pressure now acting on the ultra-thin sample with significantly smaller cross-sectional area (see also discussion below).

Batch-to-batch variations may arise from subtle differences in synthesis or annealing conditions, as well as from varying levels of residual stress or the presence of microcracks. For the example of flux-grown EuB$_6$, these issues are discussed, e.g., in Refs.\ \onlinecite{Aronson1999,Rosa2018}. Such factors can reduce the macroscopic length-change response without necessarily affecting the transition temperature. Additionally, differences in the elastic properties of the samples can influence the mechanical coupling to the dilatometer. Even within nominally identical compounds, subtle variations in composition, charge carrier concentration or structural disorder can alter the coupling strength between spin, charge and lattice degrees of freedom, potentially leading to weaker anomalies at unchanged transition temperatures. The observed reduction in signal amplitude by 20\% lies well within the range expected for such systematic effects \cite{Dagotto2005}.

\begin{figure}
\includegraphics[width=0.98\linewidth]{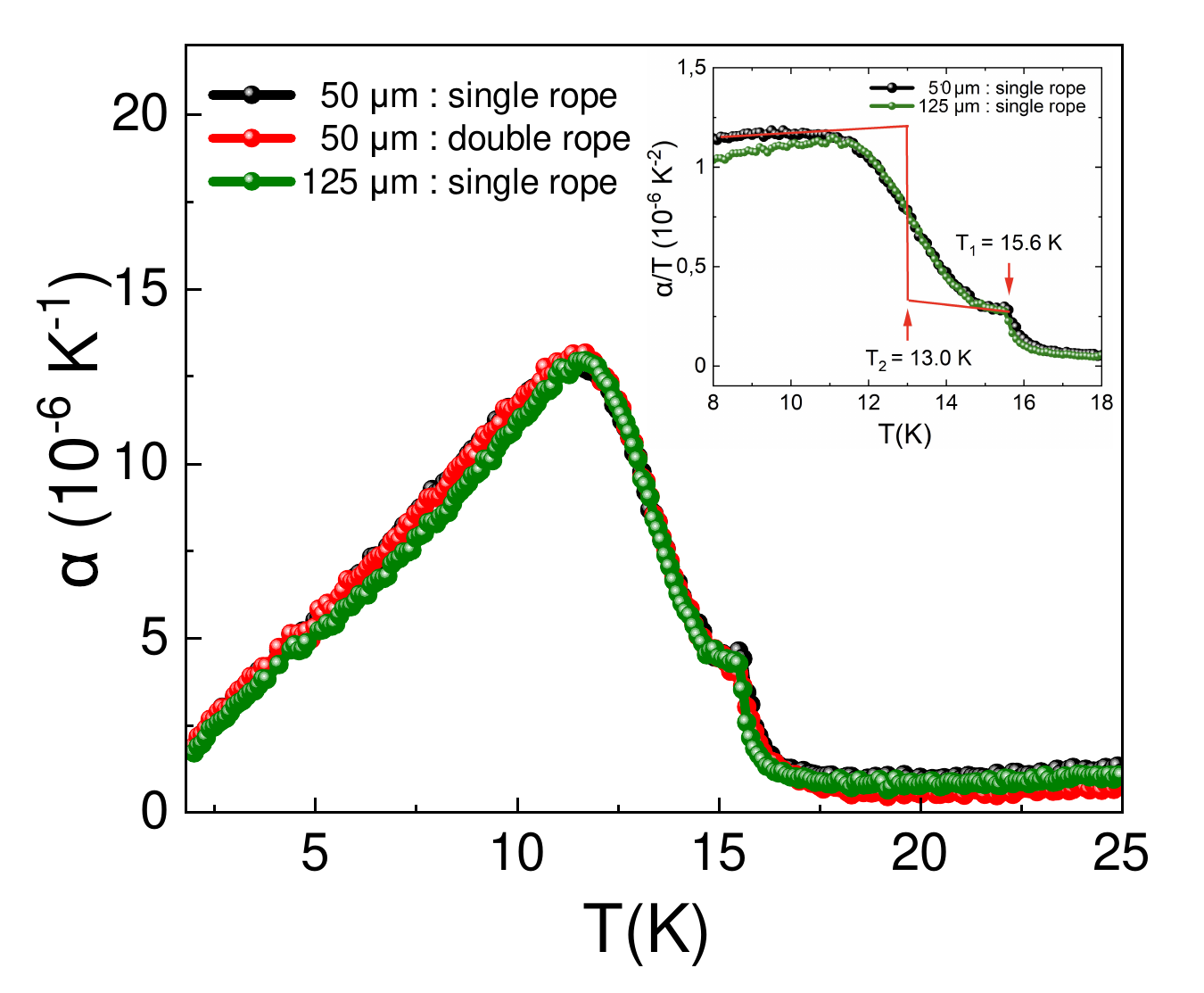}
\caption{Thermal expansion coefficient $\alpha(T)$ for the same 50~$\mu$m ultra-thin EuB$_6$ single crystal from batch \#2 mounted in the slot of a mounting stamp using
two different configurations: i) a single rope (black) and ii) a double rope for enhanced mechanical stability (red). Additionally, measurements were performed on a 125~$\mu$m thin crystal from the same batch \#2 mounted with a single rope. All measurements show good agreement. Inset shows the same data for the 50~$\mu$m crystal  in a plot $\alpha(T)/T$ vs. $T$, where the two distinct anomalies at $T_1$ = 15.6 K and $T_2$ = 13.0 K were identified by a sharp transition and an entropy-conserving equal-area construction, respectively.}
 \label{Fig14}
\end{figure}

Upon mounting for thermal expansion measurements, the 50~$\mu$m-thick sample was clamped using an adjusting screw, resulting in a measurement capacitance of 20 pF. This corresponds to a spring force of 4 N  \cite{Kuechler2023} over the sample’s cross-sectional area of approximately (0.050 × 1) mm$^2$, yielding a uni-axial pressure of $p=$ 80~MPa (0.8 kbar). Such stress is non-negligible and must be carefully considered in the analysis of thermal expansion data, as it may alter the magnitude of thermal expansion anomalies and/or slightly shift transition temperatures. Applying Ehrenfest's relation for the pressure dependence of second-order phase transitions, $(dT_C/dp)_{p \rightarrow 0}=V_{mol}T_C (\Delta \alpha/\Delta C)$, where $V_{mol}$ is the molar volume and $\Delta \alpha$ and $\Delta C$ represent the discontinuities in thermal expansion and specific heat, respectively, Manna \textit{et al}.\cite{Manna2014,Manna2014a} used literature values for $\Delta C$ to estimate the pressure derivatives of the transition temperatures. They obtained $(dT_C/dp)_{p \rightarrow 0}$ = 0.35 K/kbar for $T_1$ and 0.22 K/kbar for $T_2$, respectively, which is in good agreement with hydrostatic pressure experiments\cite{Manna2014a}. Using these coefficients and the estimated stress of 0.8 kbar exerted by the dilatometer setup, we anticipate shifts of the transition temperatures of $\Delta T_1 = 0.28$ K and $\Delta T_2 = 0.176$ K, respectively. This results in values of $T_1 = 15.58$ K and
$T_2 = 12.976$ K at $p=$ 0.8 kbar. As shown in the inset of Fig.\ \ref{Fig14}, the observed transition temperatures are in excellent agreement with these estimates, appearing at
$T_1 \approx 15.6$ K and $T_2 \approx 13.0$ K.

Although the uniaxial pressure exerted on the sample by the dilatometer setup is comparatively small and hence, the observed shifts in the transition temperatures small, they are clearly resolvable within the experimental resolution. Notably, the transition at $T_2 =$ 13.0~K exhibits considerable broadening, which complicates precise determination of the transition temperature and results in an estimated uncertainty of at least $\pm$ 0.1~K. A more rigorous
validation of the predictive capability of the Ehrenfest relation would require an extended set of measurements on samples with well-defined and systematically varied cross-sections. Such an investigation, however, lies beyond the scope of the present study. Nonetheless, one may speculate that, for materials of sufficient mechanical stability, the application of higher uniaxial pressure may provide additional insight into the material's properties. Here we note
that our recently developed miniaturized stress dilatometer\cite{Kuechler2023}, capable of applying forces of up to 65~N to a sample, might be utilized for this purpose as it employs
the same sample mounting mechanism as the one used in this study and hence, could also be equipped with the slotted stamp of Fig.\ \ref{Fig2}B.

We now address the absolute value of $\Delta L(T)/L_0$ of the 50~$\mu$m-thick sample, which is reduced by about 20\% compared to those of the thicker samples, \textit{cf.} Fig.\
\ref{Fig12} and Figs.\ \ref{Fig8},\ref{Fig9}. In order to differentiate between a batch-to-batch variation or an impact of the increased applied pressure, an additional sample from batch \#2 (from which also the 50~$\mu$m-thick sample was taken) with a
larger thickness of 125~$\mu$m, and hence a larger cross-sectional area, was investigated.
A comparison of the results is presented in Fig.\ \ref{Fig14}. Clearly, the thermal expansion measurements of both samples agree very nicely, with consistent absolute values.
This confirms that the variation in signal amplitude arises from minute differences in material properties between the two sample batches.

A close examination of the $\alpha(T)/T$-data in the inset of Fig.\ \ref{Fig14} reveals that the two transition temperatures of the 125~$\mu$m-thick sample are slightly shifted
to lower values. This shift can consistently be interpreted within the framework of Ehrenfest’s relation, taking into account the larger sample cross-section and the resulting reduction in
uniaxial pressure.

\begin{figure}
\includegraphics[width=0.98\linewidth]{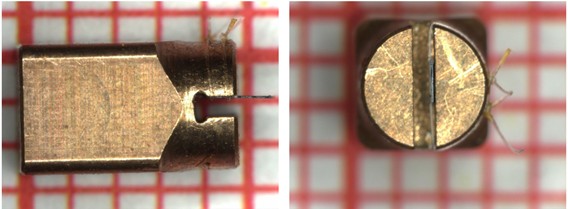}
 \caption{ A 50 $\mu$m EuB$_6$ ultra-thin crystal  with parallel edges is mounted on the narrow slot using double-knot fixation. Left picture: Side view of the mounted crystal. Right picture: Top view of the EuB$_6$ crystal mounted on the narrow slot.  }
 \label{Fig13}
\end{figure}

Misalignment of mounted samples within the slot can result in mechanical stress and potential breakage during clamping inside the dilatometer. To reduce this risk, particularly
when handling ultra-thin crystals, a second thread was introduced to enhance mechanical stability. This approach was employed in securing the 50~$\mu$m-thick EuB$_6$ single crystal, as illustrated in Fig.\ \ref{Fig13}. To verify the stability of the mounting, the measurement
was repeated following the double-thread fixation, with no detectable changes observed in the
results, see red curve in Fig.\ \ref{Fig14}. It should be noted that introducing a second
thread may also help in reducing the risk of bending the sample during the measurement.

\subsection{\label{chap3c} Test measurement on 40~$\mu$m thin AgCrS$_2$ platelet}

In the Introduction, we described a thermal-expansion measurement on a 10~$\mu$m-thin AgCrS$_2$ platelet, where the expansion was measured perpendicular to the layers, that is, along the thin (10~$\mu$m) dimension of the sample. In contrast, the measurement presented in this chapter was performed on a substantially thicker AgCrS$_2$ single crystal (40~$\mu$m), with the thermal expansion probed perpendicular to this thin dimension. This geometry allowed us to investigate a significantly larger sample, with a measurable length of $L_0$ = 1.38 mm. Inset B of Fig. 1 shows the thermal expansion coefficient $\alpha(T)$ of the very thin 10~$\mu$m sample, obtained after subtraction of the empty-cell background. As discussed previously, the pronounced noise level arises from the  $1/L_0$ scaling of $\alpha(T)$, which magnifies the impact of even very small length fluctuations. For a sample only 10~$\mu$m along the measurement axis, the intrinsic signal becomes comparable to the instrumental noise, leading to a substantial increase in the data noise. Despite these limitations, the structural anomaly at 36.5 K remains discernible on a qualitative level.

\begin{figure}
\includegraphics[width=0.6\linewidth]{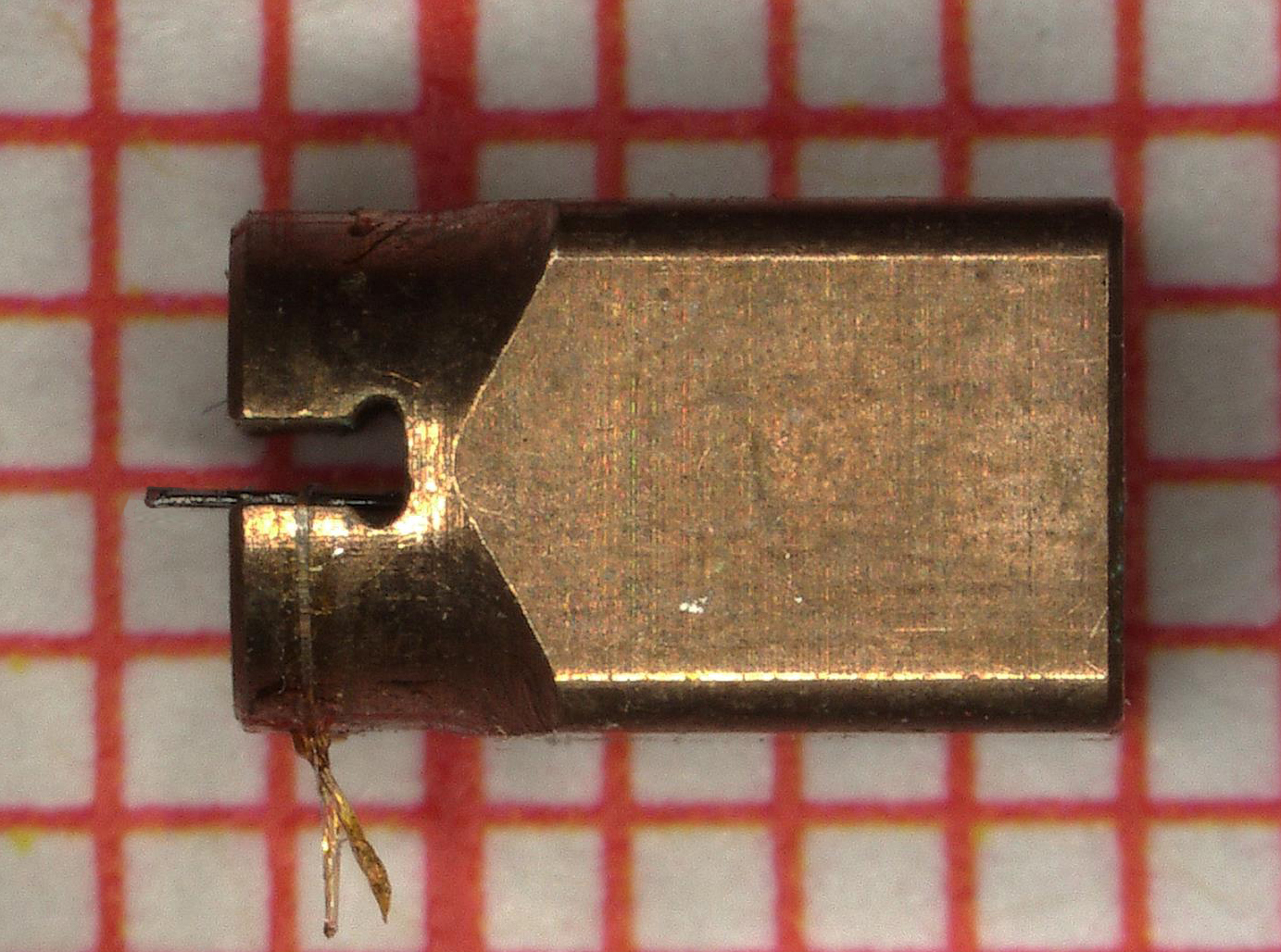}
 \caption{Side view of a 40~$\mu$m-thick AgCrS$_2$ ultra-thin single crystal mounted inside the narrow slot of the cubic stamp. The length of the sample is $L_0$ = 1.38 mm.}
 \label{Fig15x}
\end{figure}

\begin{figure}
\includegraphics[width=0.99\linewidth]{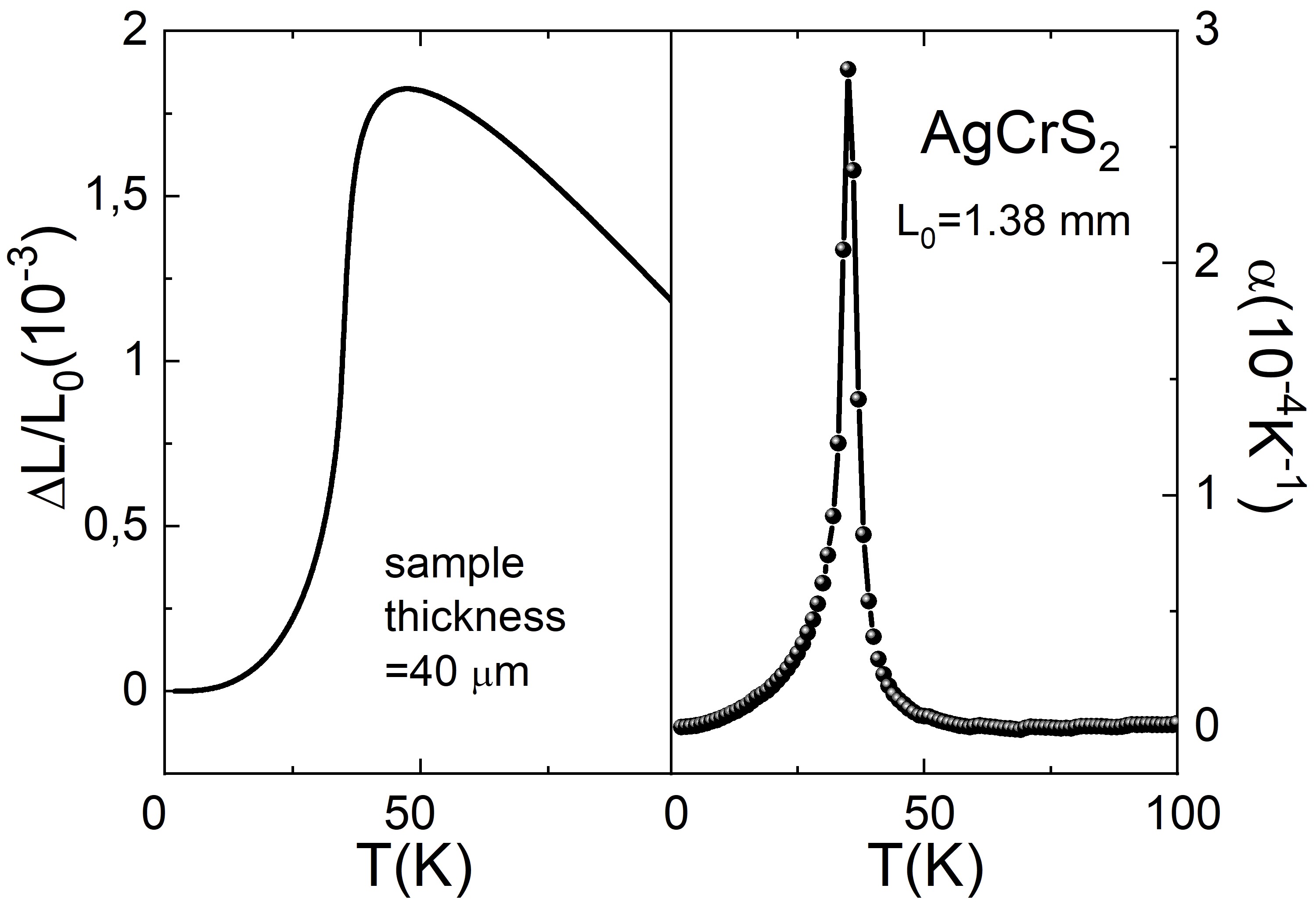}
 \caption{Relative length change $\Delta L(T)/L_0$ and  thermal expansion coefficient $\alpha(T)$ for a $L_0$ = 1.38 mm large single crystal mounted inside the slot of the custom-made cubic mounting stamp. The thickness of the measured sample is 40~$\mu$m.}
 \label{Fig16x}
\end{figure}

Recently, improved crystal-growth procedures at our institute have produced substantially thicker AgCrS$_2$ single crystals. This enabled us to prepare a 40~$\mu$m-thick sample that could be mounted inside the narrow slot of the cubic stamp (see Fig.\ \ref{Fig15x}. Importantly, the measurement could now be performed along the larger in-plane axis of the crystal ($L_0$ = 1.38 mm), which dramatically improves the signal-to-noise ratio. Fig.\ \ref{Fig16x} presents the resulting high-resolution data: the left panel shows the relative length change $\Delta L(T)/L_0$, and the right panel the corresponding thermal expansion coefficient $\alpha(T)$, both recorded on heating. A sharp, step-like anomaly in $\Delta L(T)/L_0$ is observed at T $\sim$ 36 K, corresponding to a pronounced $\lambda$-type peak in $\alpha(T)$. The magnitude of this anomaly agrees well with the earlier measurement on the 10~$\mu$m platelet, but the strongly increased sample length now provides vastly enhanced resolution across the entire temperature range. The $\lambda$-anomaly exhibits a moderate initial rise, followed by an extremely steep increase to a narrow peak at the transition, and an equally sharp decrease on warming above the transition. All these features are now clearly resolved, enabling, for the first time, a precise and quantitative determination of the thermal-expansion coefficient of AgCrS$_2$. The very narrow peak in the $\alpha(T)$ anomaly demonstrates the high crystalline quality of the newly grown sample and further confirms that our new mounting setup allows the sample to reach thermodynamic equilibrium rapidly.

AgCrS$_2$ is a layered chalcogenide whose mechanical properties are largely unreported. In our handling of 40~$\mu$m-thin single crystals, AgCrS$_2$ is extremely soft and mechanically delicate. Crystals can be cut with a razor blade, readily cleave, and are easily crushed during mounting. This qualitative observation is supported by Vickers hardness measurements performed at our institute using a Vickers microhardness tester (MHT -10). The measured hardness of 0.42 GPa confirms the extremely low mechanical strength of AgCrS$_2$. This value is comparable to that of soft metals such as high-purity copper (0.4 - 0.6 GPa) and approaches the range reported for high-purity aluminum (0.15 - 0.3 GPa)\cite{CRC_Handbook}. Given the exceptional softness and pronounced elastic anisotropy of AgCrS$_2$, careful experimental handling is essential. To reduce the risk of sample damage, the dilatometer was operated at a lowered measurement capacitance of 7 pF (rather than the standard 20 pF). Although this adjustment modestly decreases the measurement resolution, it simultaneously reduces the force applied by the tensioned springs from 4.0 N to 2.85 N, improving the mechanical stability of the fragile platelets during installation and measurement.

For a platelet thickness of 40~$\mu$m, however, even this reduced spring force produces a non-negligible uniaxial stress. A force of 2.85 N acting on a (0.040 × 1.43) mm$^2$ cross-section corresponds to a uniaxial pressure of $\sim$ 47 MPa (0.47 kbar). As a result, the thermal expansion measurements were performed under finite uniaxial stress, which may cause minor changes in the anomaly amplitudes and slight shifts in the transition temperature. These effects were discussed in detail in Sec. III.B.

\section{\label{chap4}CONCLUSION}

We report a newly developed mounting technique that enables reliable dilatometry measurements
on ultra-thin samples, including platelets with thicknesses down to 40~$\mu$m. Such micrometer-thick
samples are inherently fragile and prone to bending, tilting, or misalignment when mounted without
mechanical support, which can introduce measurement artifacts and reduce reproducibility. In our
method, the sample is inserted into a precision-machined slot within the mounting stamp. This slot
acts as a guide to maintain the sample in a fixed position and orientation, ensuring
alignment along the measurement axis and minimizing angular errors, slippage, and tilting.
If fabricated with minimal clearance, the slot gently supports the sample without restricting
its thermal expansion, allowing free longitudinal movement. This configuration stabilizes
the sample mechanically and enhances measurement reproducibility.

We have demonstrated the effectiveness of this slot-mounted design using thin, platelet-like single crystals. After initial validation with the relatively hard compound EuB$_6$, we further tested the method on a mechanically fragile 40~$\mu$m-thick AgCrS$_2$ platelet. Despite its softness, pronounced elastic anisotropy, and increased brittleness, the experiment produced reliable and reproducible data. These results highlight the robustness and practicality of the slot-mounted approach. Beyond moderately hard crystals, the method can be extended to softer or more brittle materials, although such applications require careful, case-by-case evaluation, as suitability depends on hardness, elastic modulus, and fracture resistance. Overall, our findings establish a clear proof of principle that high-quality dilatometry data can be obtained from extremely thin, plate-shaped samples using this technique.

\begin{acknowledgments}
We thank the staff of the institute workshop at the MPI for CPfS in Dresden for the precise fabrication of the mini-dilatometer and the newly developed slotted cube-shaped stamp. We are also extremely grateful to Zachary Fisk (EuB$_6$), Marcus Schmidt and Vicky Hasse (AgCrS$_2$) for providing high-quality single crystals. We gratefully acknowledge Ulrich Burkhardt for conducting the Vickers hardness measurements on AgCrS$_2$. The work in Augsburg was funded by the Deutsche Forschungsgemeinschaft (DFG, German Research Foundation: TRR 360-492547816). S.P. acknowledges the financial support of the Alexander von Humboldt Foundation. 3D visualizations of Fig. 2 and Fig. 3 created by www.archlab.de.\end{acknowledgments}

\section*{Data Availability Statement}

The data supporting the findings of this study are available in this article.

\end{document}